\documentclass[twocolumn,amsmath,amssymb,amsfonts,longbibliography]{revtex4-1}

\usepackage{graphicx}
\usepackage{dcolumn}
\usepackage{bm}
\usepackage{latexsym,color}

\usepackage{braket}

\usepackage{hyperref}
\usepackage{natbib}

\hypersetup{
  colorlinks,
  citecolor=blue,
  linkcolor=blue,
  urlcolor=blue}

\newcommand{\be}{\begin{equation}}
\newcommand{\ee}{\end{equation}}

\newcommand{\ba}{\begin{eqnarray}}
\newcommand{\ea}{\end{eqnarray}}

\begin{document}

\title{Density matrix approach to photon-assisted tunneling\\ in the transfer Hamiltonian formalism}
\author{Paul S. Davids$^{*}$}
\author{Joshua Shank}

\affiliation{%
Sandia National Laboratories, Albuquerque, NM 87185, USA
}%
\date{\today}
\email{pdavids@sandia.gov}

\begin{abstract}
The transfer Hamiltonian tunneling current is derived in a time-dependent density matrix formulation and is used to examine photon-assisted tunneling.  Bardeen's tunneling expression arises as the result of first order perturbation theory in a mean-field expansion of the density matrix.  Photon-assisted tunneling from confined electromagnetic fields in the forbidden tunnel barrier region occurs due to time-varying polarization and wavefunction overlap in the gap which  leads to a non-zero tunneling current in asymmetric device structures,  even in an unbiased state.  The photon energy  is seen to act as an effective temperature dependent bias in a uniform barrier asymmetric tunneling example problem.  Higher order terms in the density matrix expansion give rise to multi-photon enhanced tunneling currents that can be considered an extension of non-linear optics where the non-linear conductance plays a similar role as the non-linear susceptibilities in the continuity equations.
\end{abstract}


\maketitle

\section{Introduction}

Tunneling is a quantum phenomena that has been extensively studied mainly using semiclassical approaches based on the Wentzel Kramers Brillioun (WKB) approximation\cite{Murphy_Good,dunham_wkb}. Bardeen introduced a many particle transfer Hamiltonian \cite{Bardeen_1961} approach to examine tunneling currents between superconductors and normal metals through an insulating gap.  The model treated the tunneling heuristically  to match the observation of gaps in quasiparticle tunneling  seen in the superconductor to superconductor\cite{giaever1960electron} and superconductor to normal metal\cite{PhysRevLett.5.147} tunneling current voltage data of Giaever.  A more  detailed transfer Hamiltonian  approach based on the second quantized operator method and canonical transformations of the operators was subsequently developed  \cite{Bardeen_1962,Cohen_1962}.  The extension of Bardeen's transfer Hamiltonian approach to include tunneling between semiconducting half-spaces  forms the basis for the independent particle  or mean-field type approach\cite{Solid_state_theory,harrison}.   This approach has been applied to a wide range of tunneling phenomena in physics from superconducting tunneling\cite{Bardeen_1961,Bardeen_1962,Cohen_1962,exp_photon_tunneli,Tien_Gordon_multiphoton,RevModPhys.57.1055} to  semiconductor devices\cite{ct_sah,RevModPhys.46.237,FREEMAN,Kane,sze2006physics,weisbuch2014quantum} to scanning tunneling microscopy\cite{Caroli_direct,chen_stm_tunneling}.

The observation of photon enhanced tunneling in superconducting tunnel junctions under microwave illumination led to  theoretical investigations  into dynamic tunneling phenomena\cite{exp_photon_tunneli}.  Tien-Gordon proposed a theory based on time-dependent interaction of the microwave photons in the tunnel barrier of a superconducting tunnel junction\cite{exp_photon_tunneli} where they applied the transfer Hamiltonian to the multi-photon tunneling processes in the insulator.  This multi-photon enhanced tunneling process in superconducting tunnel junctions results in plateaus in the current  voltage characteristics in these devices and has been used as a sensitive direct detector for millimeter waves\cite{RevModPhys.57.1055}.   Direct detection based on rectification of displacement currents in tunneling devices have been proposed and examined theoretically\cite{Sanchez_rectenna}.

Photon-assisted tunneling has been experimentally observed in antenna-coupled semiconductor superlattice devices\cite{PhysRevLett.75.4098}, and quantum dots \cite{PhysRevLett.73.3443,PhysRevB.50.2019}. These mesoscopic systems show evidence of multi-photon assisted tunneling in the current voltage characteristics under high frequency microwave illumination at cryogenic temperatures.   Modified Tien-Gordon models have been developed and well describe the data \cite{Tien_Gordon_multiphoton}.   Furthermore, Boson-assisted tunneling \cite{PhysRevLett.72.3401} in single charged defect point-contact barriers and quantum dots have shown the  Kondo effect and resonant tunneling in these structures \cite{PhysRevLett.76.1715,PhysRevB.54.16820}.  These  results have been modeled using the Anderson model for a single charged or magnetic impurity, and the tunneling currents computed using Bardeen's transfer Hamiltonian.  These new assisted tunneling observations have led to the development of advanced modeling methods, including perturbative expansions and  non-equilibrium Green's function techniques\cite{PLATERO20041} which are closely related techniques to our derived density matrix expansion at finite temperature.   Recently, photon-assisted tunneling in resonant quantum dot systems has been examine as a means of thermoelectric energy conversion\cite{PhysRevB.87.075312,TE_QD} similar to enhanced tunneling in nanoantenna-coupled MOS tunnel diodes\cite{davids2015infrared,kadlec_prapplied}.

Another key application  of tunneling is in examination of leakage currents and gate control in metal-oxide-semiconductor (MOS) device structures owing to its technological importance. The starting point for tunneling calculations is the Bardeen transfer tunnel current expression in the static approximation and the barrier transmission expression computed within the WKB approximation.   In the WKB approach, the semi-classical barrier transmission expression depends on the energy of the carriers on the side under bias and is computed for electron or hole distributions driven out of equilibrium by a bias voltage applied to the device.   Parabolic bands in the effective mass approximation are typically assumed for the metal and semiconductors and the transverse momentum integration is performed to give the  usual supply function expression for the tunneling current\cite{ct_sah,sze2006physics}.  The asymmetirc effective masses in the metal and semiconductor give rise to large current asymmetries under bias.   Tunneling calculations for non-analytic barriers can be calculated using a transfer matrix approach\cite{Ando}. 

Recently, photon enhanced tunneling in nanoantenna-coupled MOS tunnel diodes have been  studied as a new infrared direct conversion devices\cite{davids2015infrared,kadlec_prapplied}  These large area distributed tunnel diodes concentrate infrared time-varying transverse electromagnetic fields in tunneling barriers and have been shown to lead to measurable tunneling currents and generation of electrical power from a thermal source. In this paper, we examine the time-dependent transfer Hamiltonian in a mean-field density matrix perturbative expansion.   The transfer Hamiltonian problem is expanded into complete and orthogonal bases on the right and left half-spaces, which are extended over the entire domain but not complete\cite{Prange,Appelbaum_Many_body_1969}.   The Bardeen tunneling current formula is found as the first order expression in the mean-field current expansion.  Higher order terms in the density matrix expansion are seen to give rise to resonant and non-resonant multi-photon enhanced tunneling currents.  A uniform barrier tunneling problem is examined under incoherent infrared illumination between dissimilar materials and single-photon enhanced tunneling currents are calculated for the asymmetric tunneling problem.

\section{Transfer Hamiltonian approach}
The transfer Hamiltonian approach to tunneling was originally formulated by Bardeen as a semi-empirical approach to calculate tunneling currents between a superconductor and a normal metal state through a very small gap.  Since then it has been applied to a host of general tunneling problems.    The transfer Hamiltonian approach is formulated as two half-spaces separated by a thin barrier interaction that connects the two half-spaces such that we can describe the system phenomenologically as
\be
H = H_l + H_r + H_T(t),
\label{eq:transferH}
\ee
where r and l refer to right and left half-spaces and $H_T(t) = \lambda \mathbf{V}\cos(\Omega t)$ is the transfer Hamiltonian.    The splitting of the problem into left right half-spaces imposes conditions on the completeness of the states.  The eigenstates on the left and right should be complete on their respective support and exist in the entire range of the total Hamiltonian.  This requirement is not compatible with the requirement that the left and right Hamiltonian commute. We therefore give up on the notion that the left and right energy is observable and the states of the unperturbed Hamiltonian are not left-right product states \cite{Prange}.  

With these caveats in mind, the eigenvalue problem in the two half spaces can be readily stated, for the left half space $x<0$, we have $\ket{\Psi_l }= \sum_k a_k(t) \ket{\phi_k}$ such that
\be
H_l \ket{\phi_k} =E_k^l  \ket{\phi_k}
\ee
and we have the half-space orthogonality $\braket{\phi_{k'}  | \phi_k}_l = \delta_{k,k'}$.  Similiarly, the right half space $x\ge t_{ox}$ , we have $\ket{\Psi_r} = \sum_q b_q(t) \ket{\psi_q}$ with 
\be
H_r \ket{\psi_q} = E_q^r \ket{\psi_q} ,
\ee
and $\braket{\psi_{q'}| \psi_q}_r=\delta_{q,q'}$. The left and right eigenvectors are orthogonal on their respective half-spaces but extend into the gap region, $0 < x \leq t_{ox}$ and beyond.  
Many particle Green's function approaches  have been used to examined tunneling in the transfer Hamiltonian approach\cite{Appelbaum_Many_body_1969,PLATERO20041}.  Density matrix approaches are closely related to these Green's function approaches but give simplified time-dependent perturbation expansions and are amenable to mean-field approximations at finite temperature.  

In the following sections, we will show that the Bardeen expression for the tunnel current corresponds to a first order term in a systematic density matrix expansion in the strength of the interaction.  This expansion in the density matrix will lead to a generalized non-linear current analogous to the non-linear polarization from non-linear optics.  These higher order terms correspond to resonant and non-resonant enhanced multi-photon tunneling currents. The transfer interaction will be considered  to arise from the electromagnetic coupling term confined with in the tunnel barrier and we will derive the photon-assisted tunneling current induced from a broadband incoherent source in a simple uniform barrier model.  

\section{Multi-photon density matrix expansion}
The transfer interaction for a confined electromagnetic field in the gap can be examined by considering a semi-classical electromagnetic field interaction that  takes on the familiar form,
\be
H_T(t) = i\hbar\frac{e}{mc} \mathbf{A}(t)\cdot \nabla ,
\label{eq:barrier_int}
\ee
where $\mathbf{A}(t)$ is the spatially-varying real vector potential in the gap region.  The transfer interaction from a broadband source can be expressed in terms of the  time-dependent electric field,
\be
H_T(t) =  \int_{-\infty}^{\infty} \frac{d\omega}{2\pi} e^{-i\omega t} \frac{e\hbar}{m \omega} \mathbf{E}(\omega) \cdot \nabla,
\label{eq:transfer}
\ee
where we have used the Fourier transform of the vector potential  to derive the electric field along with the constraints of real valued fields. 
This transfer interaction is shown  to transfer a quasi-particle from right to left and and vice versa, and for the above form of the  interaction is driven be the induced polarization from the electromagnetic field in the gap.    The states of the system are not product states, but single particle states that are defined by projection specified with the general index $n$.  The two non-interacting half-spaces satisfy
\be
H_0 \ket{n} = E_n \ket{n},
\ee
where $H_0 = H_l + H_r$ , where  $E_n = \{E_{n_r} , E_{n_l}\}$ depending on whether projected into right or left half-space.  

The density matrix operator can be defined in terms of the single particle states and is given by
\be
\rho =\sum_{n,m} \rho_{nm} \ket{n} \bra{m}.
\ee
The Liouville equation gives the equation of motion for the density matrix is 
\begin{widetext}
\be
i\hbar \frac{\partial \rho_{nm} }{\partial t}  =    (E_n -E_m)\rho_{nm} +  \sum_{n'm'} \big( \braket{n|H_T n'} \delta_{m,m'}  -  \braket{m'H^*_T|m}\delta_{n,n'} \big) \rho_{n'm'}.
\ee
\end{widetext}
where $\braket{m|H_T n}$ means that $H_T$ operates on the nth state. (see Appendix \ref{GDM} for general derivation and Fourier analysis of the electromagnetic transfer interaction). The Liouville equation for the density matrix can be solved using a formal perturbation expansion in the strength of the interaction, $H_T(t) \rightarrow  \lambda H_T(t))$, where $\lambda$ is  treated as a formal parameter in perturbation. Here the density matrix expansion is expanded in the interaction strength to give 
\be 
\rho_{nm} = \rho^{(0)}_{nm}  + \lambda \rho^{(1)}_{nm}(t) + \lambda^2 \rho^{(2)}_{nm}(t)+ \lambda^3 \rho^{(3)}_{nm}(t) \ldots,
\label{eq:rho_expand}
\ee
where the unperturbed density matrix is $\rho^{(0)}_{nm} = f_n \delta_{nm}$  and is assumed to be time-independent and diagonal and $f_n$ is the static non-equilibrium distribution function. 
 
We can perform the same Fourier analysis for the time-dependent density matrix and is derived in Appendix \ref{AppB}, to give
\be
\rho^{(i)}_{nm}(t) = \int\limits_{-\infty}^{\infty} \frac{d\omega}{2\pi} e^{-i\omega t}~ \tilde{\rho}^{(i)}_{nm}(\omega).
\ee
By grouping the terms in the strength of the interaction, we obtain the series expansion terms in the freqency domain.  We have the density matrix expansion terms 
\begin{widetext}
\ba
\tilde{\rho}^{(0)}_{nm}(\omega) & = & f_n \delta_{n,m} \delta(\omega ) \\ \nonumber
\tilde{\rho}^{(1)}_{nm}(\omega) & = &G_{nm} (\omega)\sum_{n'm'}  \int\limits_{-\infty}^{\infty} \frac{d\omega '}{2\pi} \tilde{T}_{nm;n'm'}(\omega - \omega ') \tilde{\rho}^{(0)}_{n'm'}(\omega ') \\ \nonumber
\tilde{\rho}^{(2)}_{nm}(\omega) & = &G_{nm} (\omega)\sum_{n'm'}  \int\limits_{-\infty}^{\infty} \frac{d\omega '}{2\pi} \tilde{T}_{nm;n'm'}(\omega - \omega ') \tilde{\rho}^{(1)}_{n'm'}(\omega ') \\  \nonumber
\tilde{\rho}^{(3)}_{nm}(\omega) & = &G_{nm}(\omega)\sum_{n'm'}  \int\limits_{-\infty}^{\infty} \frac{d\omega '}{2\pi} \tilde{T}_{nm;n'm'}(\omega - \omega ') \tilde{\rho}^{(2)}_{n'm'}(\omega ') \\  \nonumber
~~~~~~~\vdots  & & ~~~~~~~~~~~~~~~~~\vdots  \nonumber
\label{eq:dens_exp}
\ea
\end{widetext}
where  $G_{nm}(\omega) =(\hbar \omega -E_n + E_m)^{-1}$ is the free carrier propagator, and we define
\be
\tilde{T}_{nm;n'm'} (\omega)=   \big( V_{n,n'} \delta_{m,m'}  -  V_{m',m}\delta_{n,n'} \big),
\ee 
and 
\be
V_{n,n'} (\omega)=  \frac{e\hbar}{m \omega} \int d\mathbf{x} ~\mathbf{E}(\omega) \cdot \big( \phi^*_{n} (\nabla \phi_{n'} \big).
\ee 
The iterated density matrix can be expressed as an integral equation in the frequency domain and is derived  in Appendix \ref{AppB}.
The first order correction to the density matrix is 
\be
 \tilde{\rho}^{(1)}_{nm}(\omega) =G_{nm}(\omega) V_{n,m} \big(f_m - f_n \big),
 \label{eq:density1}
\ee
where we have assumed a uniform transverse electric field.

\begin{figure*}[!htb]
\centering \includegraphics[width=0.9\textwidth]{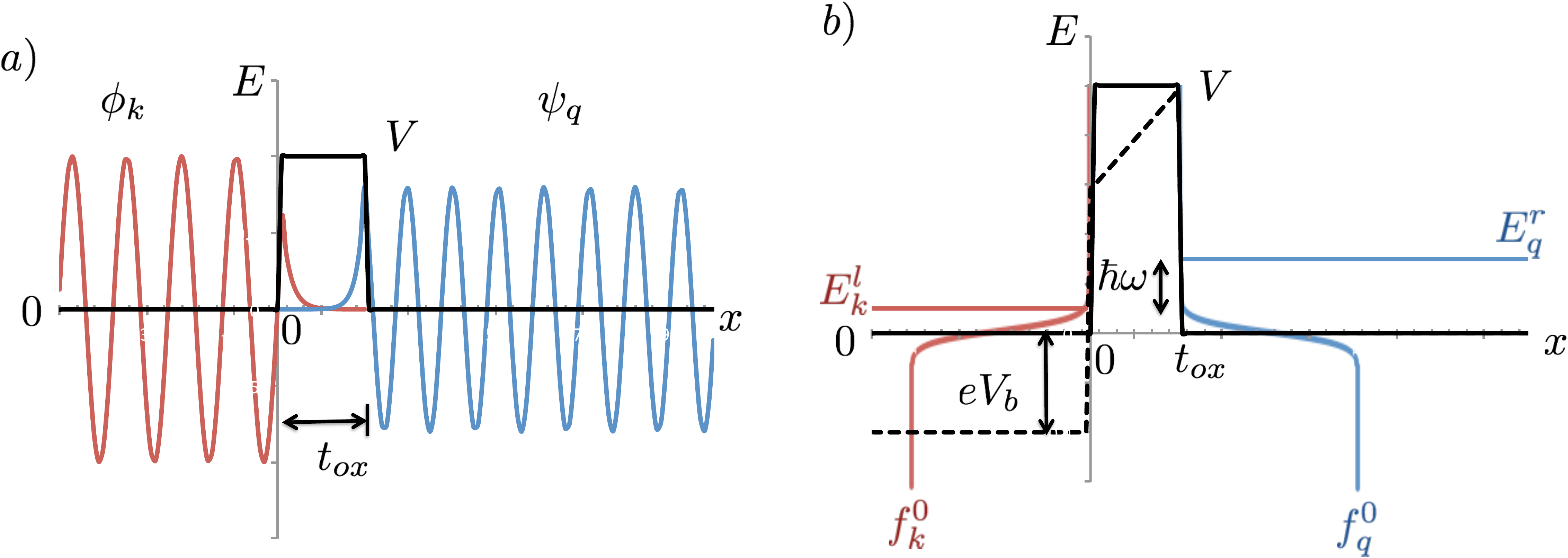}
\caption{{\bf Illustration transfer interaction for a simple 1D potential barrier.}  (a) The schematic shows the right and left wave functions, $\phi_k$ and $\psi_q$, respectively. These become evanescent in the potential barrier region.  The overlap in this region leads to the tunneling current.  (b) Energy level diagram showing the one photon tunneling process at zero bias. The quasiparticle occupancy of the left and right half-spaces are shown schematically.   The schematic impact of applying a bias (dashed black lines) leads to shift in the bands on the right side of the barrier with a spatially varying potential in the barrier.    }
\label{fig:1}
\end{figure*}

The hierarchy of density matrices in the dynamic tunneling case allows for multi-photon processes to enhance the tunneling current within the device structure.  The dynamic tunneling current  and the photon-assisted static current are derived in Appendix \ref{AppC}.  The dynamic current is given by 
\be
I(\omega')  = i\frac{2e}{\hbar} \sum_{nm}   \int\limits_{-\infty}^{\infty} \frac{d\omega}{2\pi}V_{m,n}(\omega) ~\tilde{ \rho}_{nm}(\omega'-\omega),
\ee
and the static tunneling current is 
\be
I  = -i\frac{2e}{\hbar} \sum_{nm}   \int\limits_{-\infty}^{\infty} \frac{d\omega}{2\pi}\tilde{V}_{m,n}(\omega) ~\tilde{ \rho}_{nm}(\omega).
\ee
Here we have 
\be
\tilde{V}_{m,n'} (\omega)=  \frac{e\hbar}{m \omega} \int d\mathbf{x} ~\mathbf{E}^*(\omega) \cdot \big( \phi^*_{m} (\nabla \phi_{n'} \big).
\ee 

The expansion of the density matrices in the strength of the interaction leads to a  similar expansion of the static and dynamic currents.  We have that 
\be
I = I^{(0)} + \lambda I^{(1)} + \lambda^2 I^{(2)} + \lambda^3 I^{(3)} \ldots
\ee
from the expansion of $\tilde{\rho}$.   This expression gives rise to an expansion for the non-linear conductances in the same fashion as the non-linear susceptibilities arise in non-linear optics.  Note, the static current at zero order is not included in the expansion, the static part of the density matrix is included in the dynamic zeroth order term and has the frequency dependence of the electric field only.

The first order term in the static current expression can be written explicitly,
\be
I^{(1)}  = -i\frac{2e}{\hbar} \sum_{nm}   \int\limits_{-\infty}^{\infty} \frac{d\omega}{2\pi}\tilde{V}_{m,n}V_{n,m}  G_{nm}(\omega) \big(f_m - f_n \big),
\ee
where we have used Eq. \ref{eq:density1} in obtaining the above current.    The first order current can be simplified if we consider the special case of a uniform transverse electric field in the gap and use the principal value expression for the Green's function energy denominator
\be
G_{nm}(\omega) = \text{P} \left(\frac{1}{\hbar \omega -E_n + E_m}\right)  +i\pi \delta(\hbar \omega -E_n + E_m),
\ee
to obtain,
\begin{widetext}
\be
I^{(1)}  = \frac{2\pi e}{\hbar} \int\limits_{-\infty}^{\infty} \frac{d\omega}{2\pi}  \left(\frac{e\hbar}{m \omega}\right)^2 |E_z(\omega)|^2  \sum_{nm} u_{mn} u_{nm} \big(f_m - f_n \big)\delta( \hbar \omega - E_n + E_m),
\label{eq:bigone}
\ee
\end{widetext}
 where $u_{nm} = \braket{n|\nabla_z m}$,and which represents the first order real-valued static current induced by a frequency-dependent transverse electric field.   Eq. \ref{eq:bigone} is a generalized Bardeen transfer expression for photon-induced tunneling from a broadband radiation source.   The frequency integral can be restricted to positive frequencies leading to two contributions to the photon-assisted current corresponding to  absorption and emission of a photon. 

The higher order currents can be derived by iterating the density matrix expansion in Eq. \ref{eq:dens_exp}, but lead to much more complex expressions.    The second order static current can be obtained in the uniform field case,
\begin{widetext}
 \ba
 I^{(2)} & = & -i\frac{2e}{\hbar} \int\limits_{-\infty}^{\infty} \frac{d\omega}{2\pi} \left( \frac{e\hbar}{m \omega} \right)  \int\limits_{-\infty}^{\infty} \frac{d\omega'}{2\pi} \left( \frac{e\hbar}{m \omega'}\right)  \left(\frac{e\hbar}{m (\omega-\omega')} \right) E_z^*(\omega) E_z(\omega')   E_z(\omega-\omega')  \nonumber \\
 & & ~~~~\sum_{ln'm'} u_{m'l}u_{ln'}u_{n'm'} [f_{m'} - f_{n'}]G_{n'm'}(\omega') \left( G_{m'l}(\omega) - G_{ln'}(\omega)  \right),
 \ea 
 \end{widetext}
 where $u_{nm} = \braket{n|\nabla_z m}$.   The 2nd order current is obtained from the real value of the above expression and is cubic in the electric field strength and has a complex frequency dependence.  In the following, we will consider a special example of a uniform barrier with different contact materials on either side of the barrier.   This simple problem will be evaluated using our 1st order static current result (Eq. \ref{eq:bigone}) and will show  photon-assisted tunneling in an simple example problem with technical relevance.

\section{Uniform Barrier Example}

It is instructive to examine a simple example. We consider tunneling through a uniform barrier in the presence of a transverse uniform time varying electromagnetic field. The electric field is from an incoherent black-body source with a small and finite bandwidth.   We consider a simple transfer Hamiltonian of the form,
\be
H=   \left( \frac{\mathbf{p}^2}{2m_{ox}}+V \right) + H_T(t),
\ee
where  $V$ is a real-valued constant barrier height in the region $0< x\le t_{ox}$,  and last term is the imaginary potential that arises from time-varying vector potential.  Eq. \ref{eq:bigone} can be  applied to the uniform barrier problem to evaluate the first order tunneling current. The wavefunctions in the barrier from the left and right regions are given by 
\ba
\phi_k&  = & \sqrt{p_l} \exp(i\mathbf{k}\cdot \mathbf{r} -k x)\\
\psi_q &=&  \sqrt{p_l}\exp(i\mathbf{q}\cdot \mathbf{r}-q (t_{ox} - x))
\ea
where $\mathbf{k}$ and $\mathbf{q}$ are the transverse momentum, and  $k =\sqrt{2m_l (V -E_l)/\hbar^2}$ and $q =\sqrt{2m_r (V -E_r)/\hbar^2}$, where $V$ is the barrier height. Here $p_{l,r}=1/V_{l,r}$ are the real-valued volume normalizations  in the left and right half-spaces.    Figure \ref{fig:1} (a) shows  schematically the wavefunctions in the two half-spaces and the overlap in the barrier.    The wavefunction  in the barrier region exponentially decays away from the left and right interfaces respectively. The tunneling occurs due to the dipole polarization induced by the field confined in the gap and  the wavefunction overlap in this region. 

The matrix element of the transfer interaction are readily evaluated
\ba
u_{qk} & = & - k~ T_{qk} (2\pi)^2\delta (\mathbf{q}-\mathbf{k}),
\ea
and similiarly for the transpose
\ba
u_{kq} & = &  q~T_{kq} (2\pi)^2\delta (\mathbf{k}-\mathbf{q}),
\ea
where  transverse momentum conservation is explicit, $\mathbf{k} = \mathbf{q}$, and  $T_{kq}=T_{qk}$ is symmetric with
\be
T_{qk} = \frac{t_{ox}}{\sqrt{V_l V_r}}e^{-(k+q)t_{ox}/2} \frac{ \sinh((k-q)t_{ox}/2)}{(k-q)t_{ox}/2}.
\ee
The imaginary part of the transfer interaction matrix elements occurs due to the momentum operator matrix elements with the evanescent decaying wavefunctions.   The imaginary transfer interaction is needed to give rise to a non-zero current through the barrier and implies that we are not in an equilibrium situation but a dynamic situation in which quasiparticles are injected and removed from a reservior at the left and right contacts.  

\begin{figure*}[!htb]
\centering \includegraphics[width=0.9\textwidth]{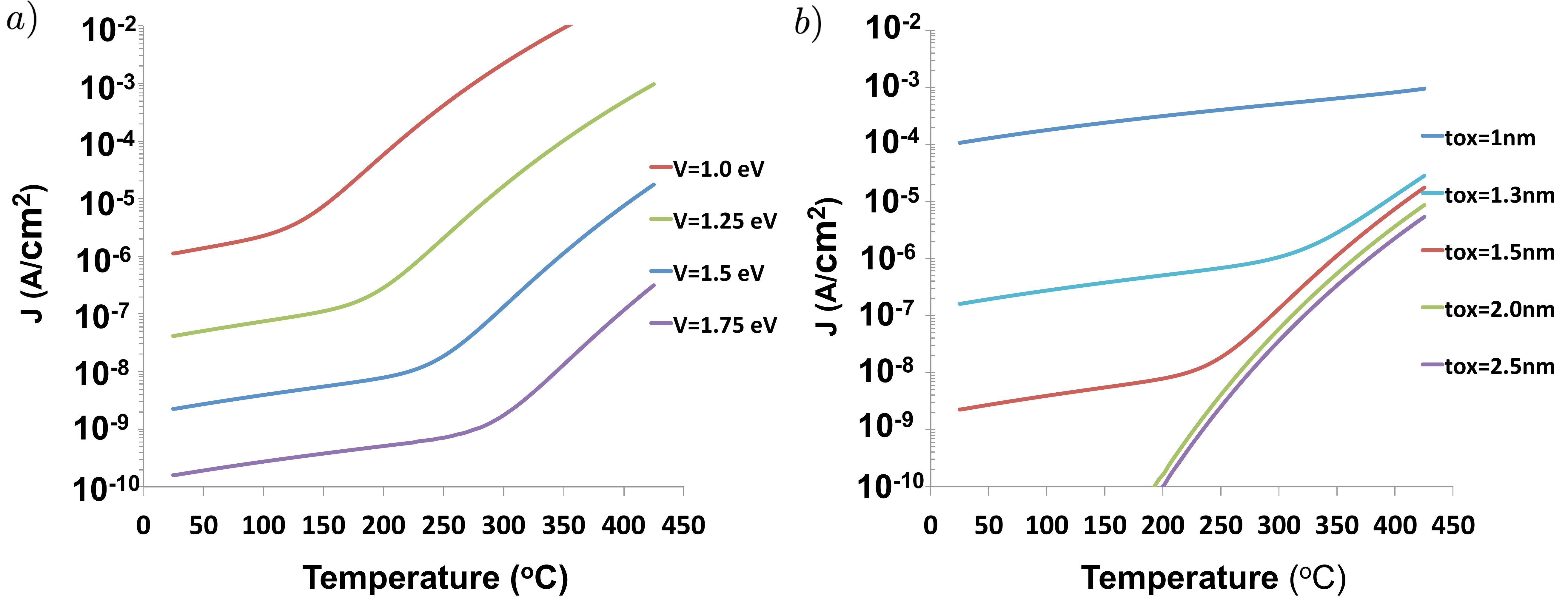}
\caption{{\bf Tunneling current density for various barrier parameters.}  (a) Computed tunneling current for varying barrier energy $V$ for fixed barrier thickness $t_{ox} =1.5$nm as a function of temperature.     (b) Computed tunneling current for different barrier thicknesses $t_{ox}$ for fixed barrier energy, $V=1.5$eV as a function of temperature.   For all calculations, $m_l=m_e$, $m_r=0.19m_e$, and $m_{ox} = 0.5m_e$ where $m_e$ is the bare electron mass. The photon energy $\hbar \omega = 0.17$ eV corresponding to a wavelength of approximately 7.3 microns.   The field in the barrier, $E_0$, is obtained from a blackbody distribution for the specified temperature at the above central wavelength with $d\nu=1$ THz bandwidth.   }
\label{fig:3}
\end{figure*}

The first order photon transfer tunneling  current is given by Eq. \ref{eq:bigone}, 
and  the sums are over both transverse and longitudinal wavevectors. These sums are converted to integrals 
\be
\sum_k \rightarrow V_l  \int \frac{dk}{2\pi} \int \frac{d\mathbf{k}}{(2\pi )^2}
\ee 
over longitudinal and transverse momenta and similarly for right half-space momenta $q$.
The current density can be obtained by integrating over the transverse momentum in the left and right half-spaces.  This can be done analytically if the parallel effective masses in the left and right half-spaces are assumed to be equal, $m_e$,  and the evanescent x wavevectors do not depend on the transverse momenta.   The single photon tunneling current is 
\begin{widetext}
\be
J^{(1)} = \frac{4\pi e m_e }{h^3\beta^2} \left( \frac{m_r m_l}{m^2_{ox}} \right) \int\limits_{0}^{\infty} \frac{d\omega}{2\pi}\left( \frac{e t_{ox}}{ \hbar \omega} \right )^2 |E(\omega)|^2  {\cal T} (\omega), 
\label{eq:current_1}
\ee
where the transmittance is defined as 
\be
{\cal T} (\omega) = \beta \int dE_l~ t(E_l,E_l+  \hbar \omega)  \log \left[  \frac{ 1+\exp(-\beta E_l)  }{1+ \exp(-\beta(E_l  +\hbar \omega)   }  \right] + (\omega \rightarrow -\omega) ,
\label{eq:trans}
\ee
\end{widetext}
and where $\beta = 1/k_B T$,  and the photon energy $ \hbar \omega$ acts as an effective bias in the supply function. There is a smaller back-flow current contribution that corresponds to photon emission in the barrier, $E_l - \hbar \omega$. This contribution to the transmittance has been ignored.   In  Eq. \ref{eq:current_1}, the first term has  the units  for the current density, $AT^2$ where $A =120 \text{ Amps}/ \text{cm}^2 \text{K}^2$ is Richardson's constant.  The second expression in parenthesis is the ratio of the x-directed perpendicular effective masses for the right and left half-spaces and oxide barrier.  The third term is  dimensionless interaction strength and depends on the amplitude of the field in the barrier, $|E|$ and the transmittance, ${\cal T} (\omega)$ .  This integral is over all positive frequencies, but can be limited to a finite bandwidth $d\nu$ near the photon energy.  The  transmittance is the scaled integral over the barrier transmittance and supply function give the overall transmittance.  The barrier transmittance is 
\be
t(E_l,E_r) =e^{-(k+q)t_{ox}}~ \text{sinhc}^2\left(\frac{(k-q)t_{ox}}{2} \right),
\ee
where the first term has the form of the standard tunneling exponential and the second term is the square hyperbolic sinc function. The evanescent wavevectors in the barrier region are  $k =(2m_l (V -E_l)/\hbar^2)^{1/2}$ and $q =(2m_r (V -E_r)/\hbar^2)^{1/2}$ as previously defined.

Figure \ref{fig:3}(a) shows the computed first order current for fixed  barrier thickness and (b) for fixed barrier height as the device temperature is varied from 25$^o$C to 425$^o$C.  A key component of the calculation is the photon field amplitude in the barrier, $|E|$.  The photon field amplitude is derived from a black-body spectral exitance, $M^0_{\nu} $, given by 
\be
\frac{d\Phi}{d\nu} = M^0_{\nu}(T) = \frac{2\pi h \nu^3}{c^2} \frac{1}{\exp(\beta h \nu)-1}
\ee
where $\Phi$ is the power per unit area, $\nu = \omega/2 \pi$ is the photon frequency, and $d\nu$ is the bandwidth.  The electric field amplitude in the barrier is 
\be
|E(\omega)|^2= 2Z_0 M^0_{\nu}(T) 
\ee
where $Z_0$ is the impedance of free-space.   In fig. \ref{fig:3} (a) we see that there is strong variation of the current density with barrier height with increasing current with decreasing barrier height as expected.  There is also a strong temperature dependence with the current density sharply increasing above a critical temperature.  This critical temperature decreases with lowered barrier and is due to the thermal occupancy of the left and right energy bands and the increase of the thermal photon electric field  amplitude in the barrier with increasing temperature.  Figure \ref{fig:3}(b) shows the dependence of the the tunneling current density on the barrier thickness.  The tunneling current increases with decreasing barrier thickness as expected.  Interestingly, the current density above a critical temperature  asymptotically approaches an common exponential growth  for increasing temperatures.  The barrier transmittance and the dimensionless field amplitude terms in Eq.\ref{eq:current_1} depend exponentially and quadratically on the barrier thickness, respectively,  and are the source of this asymptotic behavior.   Short circuit tunneling current in large area nanoantenna coupled tunnel diodes illuminated a thermal source have been reported experimentally \cite{davids2015infrared,kadlec_prapplied}.  In these devices, high field concentration in the tunnel barrier from a thermal infrared sources lead to high short currents with strong temperature dependence and may lead to new radiative thermoelectric direct conversion of thermal radiation into direct electric power.

\section{Conclusions}
We have derived the transfer Hamiltonian tunneling current using a time dependent mean-field density matrix perturbation expansion.  The density matrix is computed to all orders in perturbation theory and the generalized tunneling current is obtained.  The Bardeen tunneling current expression is obtained to first order in the density matrix expansion.  Photon-assisted tunneling arises due to confined electromagnetic fields in the tunnel barrier region.  These fields and wavefunction overlap in the barrier produce a time-varying polarization which leads to a net current flow.   This complex interaction in the barrier is a direct consequence of the imaginary momenta in the barrier and results in an imaginary transfer interaction that leads to the tunneling current.  The dependence on the photon energy in the density matrix hierarchy gives rise to photon-assisted tunneling through the barrier, where the photon energy acts as an effective bias.   

To examine the effective of photon-assisted tunneling, we examine a uniform barrier model with asymmetric effective masses in the two half-spaces.  We  consider the confined field in the tunnel gap arising from black-body irradiation consistent with the experimental observation of short-circuit tunneling current observed in infrared nanoantenna-coupled tunnel diode rectfiiers\cite{davids2015infrared,kadlec_prapplied}.  In these devices, electric field confinement in the tunnel gap results from strong photon-phonon interaction near a longitudinal optical phonon mode in the oxide tunnel barrier.  We find that the tunneling current has a temperature dependent turn on  as the source temperature is increased, and is seen to saturate at different temperatures for varying barrier thicknesses.  Multi-photon assisted tunneling arises from higher order terms of $eE_0 t_{ox}/\hbar\omega$, which is small and represents a minor contribution to the first order Bardeen tunneling current computed for the uniform barrier model considered.  In this paper, we have presented a generalized  density matrix approach to the computation of multi-photon assisted tunneling currents  and examined a simple 1D asymmetric tunneling model.   The effect of applying a bias and the role of surface states are challenging problems under consideration in photon-assisted tunneling devices and need to be solved within a consistent many-body framework.   

\acknowledgements{
Funding for this work was provided by Sandia's Laboratory Directed Research and Development
(LDRD) program. Sandia National Laboratories  is a multi-mission laboratory managed and operated
by National Technology and Engineering Solutions of Sandia, a wholly owned subsidiary of Honeywell International Inc., for the United States Department
of Energy's National Nuclear Security Administration under contract DE-NA0003525.
}

%

\appendix
\begin{widetext}
\section{Generalized density matrix}\label{GDM}

The transfer Hamiltonian under consideration explicitly in this paper is given by the electromagnetic interaction in the tunnel barrier, we have 
\be
H_T(t) = i\frac{e\hbar}{m c} \mathbf{A}(t) \cdot \nabla.
\ee
The real-valued vector potential can be written in terms of a Fourier transform 
\be
\mathbf{A}(t) = \int_{-\infty}^{\infty} \frac{d\omega}{2\pi} e^{-i\omega t} \mathbf{A}(\omega),
\ee
and we require $\mathbf{A}^*(-\omega) = \mathbf{A}(\omega)$ for real-valued fields.  The vector potential in the Coulomb gauge can be expressed in terms of the electric field, we have $\mathbf{E}(\omega) = i\omega/c \mathbf{A}(\omega)$ and we have $\mathbf{E}^*(-\omega) = \mathbf{E}(\omega)$.   The resulting transfer interaction is given by 
\be
H_T(t) =  \int_{-\infty}^{\infty} \frac{d\omega}{2\pi} e^{-i\omega t} \frac{e\hbar}{m \omega} \mathbf{E}(\omega) \cdot \nabla ,
\label{app:eq:transfer}
\ee
and we find that $H^*_T(t) = -H_T(t)$ but care must be taken regarding order of operations.  In the following, we will examine the equations of motion for the density matrix in great detail for clarity.

The density matrix can derived for the transfer interaction. The Hamiltonian for the system is  $H= H_0 + H_T(t)$, with $H_0\ket{n} = E_n\ket{n}$ where $H_0$ is the unperturbed Hamiltonian with orthonormal states,  $\braket{m|n} = \delta_{nm}$.  Schrodinger's equation of motion is 
\be
i \hbar \partial_t \ket{\Psi} = (H_0 + H_T(t))\ket{\Psi},
\label{app:Schrod}
\ee
and the total wavefunction  can be expanded in the unperturbed basis
$\ket{\Psi} = \sum_n c_n(t) \ket{n}$, to give
\be
i\hbar \dot{c}_n = E_n c_n + \sum_{n'} \braket{n| H_T n'} c_{n'},
\ee
and a similar expression for the conjugate gives
 \be
i\hbar  \dot{c}^{\dagger}_m = -E_m c^{\dagger}_m- \sum_{m'} \braket{m'H^*_T| m} c^{\dagger}_{m'}.
 \ee
The generalized density operator is 
\be
\hat{\rho} = \ket{\Psi} \bra{\Psi} = \sum_{n,m} c_n(t) c^{\dagger}_m(t) \ket{n} \bra{m},
\ee
where $\rho_{nm} = c_n(t) c^{\dagger}_m(t) $. The density matrix equation of motion is 
\be
\boxed{ i\hbar \frac{\partial \rho_{nm} }{\partial t}  =    (E_n -E_m)\rho_{nm} +  \sum_{n'm'} \big( \braket{n|H_T n'} \delta_{m,m'}  -  \braket{m'H^*_T|m}\delta_{n,n'} \big) \rho_{n'm'}.}
\ee
It is convenient to define the transfer matrix 
\be
T_{nm;n'm'} = \big( \braket{n|H_T n'} \delta_{m,m'}  -  \braket{m'H^*_T|m}\delta_{n,n'} \big). 
\ee
The matrix elements can be evaluated in the transfer matrix, we have
\be
T_{nm;n'm'} = \int d\mathbf{x} \int_{-\infty}^{\infty} \frac{d\omega}{2\pi} e^{-i\omega t} \frac{e\hbar}{m \omega} \mathbf{E}(\omega) \cdot \big( \phi^*_{n} (\nabla \phi_{n'} )\delta_{m,m'}  -   \phi^*_{m'} (\nabla \phi_{m} )\delta_{n,n'}  + \nabla (\phi^*_{m'} \phi_m) \big),
\ee
where the last term can be converted to a divergence which vanishes on the bounding surface and  a term proportional to $\nabla \cdot \mathbf{E} = 0$.   Therefore, we obtain
\be
T_{nm;n'm'} =  \int_{-\infty}^{\infty} \frac{d\omega}{2\pi} e^{-i\omega t} \frac{e\hbar}{m \omega} \int d\mathbf{x} ~\mathbf{E}(\omega) \cdot \big( \phi^*_{n} (\nabla \phi_{n'} )\delta_{m,m'}  -   \phi^*_{m'} (\nabla \phi_{m} )\delta_{n,n'} \big),
\ee 
where the equation of motion becomes
\be
\boxed{ i\hbar \frac{\partial \rho_{nm} }{\partial t}  =    (E_n -E_m)\rho_{nm} +  \sum_{n'm'}T_{nm;n'm'}(t)~ \rho_{n'm'}.}
\ee

\section{Time-dependent perturbation expansion}\label{AppB}
The expansion of the density matrix in the strength of the interaction is given by Eq. \ref{eq:rho_expand} in the text, we obtain

\ba
i\hbar \dot{\rho}^{(1)}_{nm} - (E_n-E_m)  \rho^{(1)}_{nm}  & =  & \sum_{n'm'}  \int\limits_{-\infty}^{\infty} \frac{d\omega}{2\pi} e^{-i\omega t}  T_{nm;n'm'} \rho^{(0)}_{n'm'}  \\
i\hbar \dot{\rho}^{(2)}_{nm} - (E_n-E_m)  \rho^{(2)}_{nm}  & =  & \sum_{n'm'}  \int\limits_{-\infty}^{\infty} \frac{d\omega}{2\pi} e^{-i\omega t}  T_{nm;n'm'} \rho^{(1)}_{n'm'}  \\
i\hbar \dot{\rho}^{(3)}_{nm} - (E_n-E_m)  \rho^{(3)}_{nm}  & =  & \sum_{n'm'}  \int\limits_{-\infty}^{\infty} \frac{d\omega}{2\pi} e^{-i\omega t}  T_{nm;n'm'} \rho^{(2)}_{n'm'} \\
\vdots & & \vdots  \nonumber
\ea
and we recall that the zeroth order term is time independent and diagonal.    We can perform the same Fourier analysis for the time-dependent density matrix,
\be
\rho^{(i)}_{nm}(t) = \int\limits_{-\infty}^{\infty} \frac{d\omega}{2\pi} e^{-i\omega t}~ \tilde{\rho}^{(i)}_{nm}(\omega).
\ee
The first order correction to the density matrix is 
\be
 \tilde{\rho}^{(1)}_{nm}(\omega) = \frac{1}{\hbar \omega -E_n + E_m} \left( \frac{e\hbar}{m \omega} \right)\mathbf{E}(\omega) \cdot \braket{n|\nabla m} \big(f_m - f_n \big),
\ee
where we have assumed a uniform transverse electric field.
The second order density matrix  and higher order terms are 
\ba
\tilde{\rho}^{(2)}_{nm}(\omega) & = & \frac{1}{\hbar \omega -E_n + E_m} \sum_{n'm'}  \int\limits_{-\infty}^{\infty} \frac{d\omega '}{2\pi} \tilde{T}_{nm;n'm'}(\omega - \omega ') \tilde{\rho}^{(1)}_{n'm'}(\omega ') \\
\tilde{\rho}^{(3)}_{nm}(\omega) & = & \frac{1}{\hbar \omega -E_n + E_m} \sum_{n'm'}  \int\limits_{-\infty}^{\infty} \frac{d\omega '}{2\pi} \tilde{T}_{nm;n'm'}(\omega - \omega ') \tilde{\rho}^{(2)}_{n'm'}(\omega ') \\
~~~~~~~\vdots  & & ~~~~~~~~~~~~~~~~~\vdots  \nonumber
\ea
where we note that 
\be
\tilde{T}_{nm;n'm'} (\omega)=  \frac{e\hbar}{m \omega} \int d\mathbf{x} ~\mathbf{E}(\omega) \cdot \big( \phi^*_{n} (\nabla \phi_{n'} )\delta_{m,m'}  -   \phi^*_{m'} (\nabla \phi_{m} )\delta_{n,n'} \big).
\ee 
The density expansion can be considered a consequence of an iterated integral equation for the total density matrix.  We have 
\be
\tilde{\rho}_{nm}(\omega) =  \tilde{\rho}^{(0)}_{nm}(\omega)  +\frac{1}{\hbar \omega -E_n + E_m} \sum_{n'm'}  \int\limits_{-\infty}^{\infty} \frac{d\omega '}{2\pi} \tilde{T}_{nm;n'm'}(\omega - \omega ') \tilde{\rho}_{n'm'}(\omega ') ,
\ee
which has the form of a scattering equation.  The matrix element $\tilde{T}$ acts as a self-energy term in the non-equilibrium density matrix and the above equation can be formally inverted for the non-equilibrium density matrix. 

\section{Current continuity}\label{AppC}
The transfer Hamiltonian for the system is given by
\be
H = H_0 + H_T(t),
\ee
where in general 
\be
H_0 = -\frac{\hbar^2}{2m}\nabla^2 + U,
\ee
with $U$ the spatially dependent real-valued potential.  The expansion of the wavefunction is also as given above, we have 
\be
\Psi = \sum_n c_n(t) \phi_n(\mathbf{x}).
\ee
From the equation of motion and the definition of the density matrix given in pervious Appendices,  we obtain
\be
 \sum_{n,m} \left( \frac{\partial \rho_{nm}}{\partial t} \phi^*_m\phi_n + \nabla \cdot \mathbf{j}_{nm}~\rho_{nm} \right) = \frac{2e}{\hbar} \sum_{nm}  \int\limits_{-\infty}^{\infty} \frac{d\omega}{2\pi} e^{-i\omega t} \frac{1}{ \omega} \mathbf{E}(\omega) \cdot \mathbf{j}_{nm}~\rho_{nm},
\ee
where the current is defined as 
\be
\mathbf{j}_{nm} = \frac{\hbar}{2mi} \big(\phi^*_m ( \nabla \phi_n )-(\nabla \phi^*_m ) \phi_n \big).
\ee
If we now integrate over an arbitrary volume, we can apply Green's theorem to convert to a surface integral over the bounding volume to obtain
\be
\frac{\partial N}{\partial t} + \iint dA \hat{n} \cdot \mathbf{j}  = \frac{2e}{\hbar}   \int\limits_{-\infty}^{\infty} \frac{d\omega}{2\pi} e^{-i\omega t} \frac{1}{ \omega} \int_V d\mathbf{x} ~\mathbf{E}(\omega) \cdot \mathbf{j},
\ee
where the current is $\mathbf{j} = \sum_{nm} \mathbf{j}_{nm}~\rho_{nm}$, and number of quasi-particles is
\be 
N = \sum_{nm} \rho_{nm} \int_V d\mathbf{x} ~ \phi^*_m\phi_n.
\ee 
The charge current  is obtained by multiplying by the $-e$ electronic charge and the electrical current density is $\mathbf{J} = -e \mathbf{j}$.   The resulting expression for the electrical current is 
\be
I =  \frac{2e}{\hbar}   \int\limits_{-\infty}^{\infty} \frac{d\omega}{2\pi} e^{-i\omega t} \frac{1}{ \omega} \int_V d\mathbf{x} ~\mathbf{E}(\omega) \cdot \mathbf{J} + e\frac{\partial N}{\partial t},
\label{eq:current}
\ee
where 
\be
I = -e \iint dA \hat{n} \cdot \mathbf{J} .
\ee
Eq. \ref{eq:current} is the general equation for current continuity in the presence of dissipation.   If we ignore the time-varying charge density term, we obtain
\be
I =  \frac{2e}{\hbar}   \int\limits_{-\infty}^{\infty} \frac{d\omega}{2\pi} e^{-i\omega t} \frac{1}{ \omega} \int_V d\mathbf{x} ~\mathbf{E}(\omega) \cdot \mathbf{J}.
\label{eq:simple}
\ee
The current from Eq. \ref{eq:simple} can be expressed in terms of the transfer Hamiltonian.  Explicitly, we have 
\be
I  = i\sum_{nm}  \int\limits_{-\infty}^{\infty} \frac{d\omega}{2\pi} e^{-i\omega t} \frac{e^2}{m \omega}  \int_V d\mathbf{x} ~\mathbf{E}(\omega) \cdot \big(\phi^*_m ( \nabla \phi_n )-( \nabla \phi^*_m ) \phi_n \big) \rho_{nm}.
\ee
Using the previous integration by parts, we get 
\be
I  = i\frac{2e}{\hbar} \sum_{nm}  \int\limits_{-\infty}^{\infty} \frac{d\omega}{2\pi} e^{-i\omega t} \frac{e\hbar}{m \omega}  \int_V d\mathbf{x} ~\mathbf{E}(\omega) \cdot \big(\phi^*_m ( \nabla \phi_n )\big) \rho_{nm},
\ee
or 
\be
\boxed{I(\omega')  = i\frac{2e}{\hbar} \sum_{nm}   \int\limits_{-\infty}^{\infty} \frac{d\omega}{2\pi} \frac{e\hbar}{m \omega}  \int_V d\mathbf{x} ~\mathbf{E}(\omega) \cdot \big(\phi^*_m ( \nabla \phi_n )\big)\tilde{ \rho}_{nm}(\omega'-\omega),}
\ee
and since we are mainly interested in the  DC current that results from the dynamic electromagnetic field, we evaluate at $\omega' = 0$
\be
\boxed{I_0  = -i\frac{2e}{\hbar} \sum_{nm}   \int\limits_{-\infty}^{\infty} \frac{d\omega}{2\pi} \frac{e\hbar}{m \omega}  \int_V d\mathbf{x} ~\mathbf{E}^*(\omega) \cdot \big(\phi^*_m ( \nabla \phi_n )\big)\tilde{ \rho}_{nm}(\omega),}.
\ee
\end{widetext}


\end{document}